# Dirac's equation in 1+1 D from a random walk

Michael Ibison

Institute for advanced studies at Austin

4030 Braker Lane West, suite 300, Austin, TX 78759

ibison@earthtech.org

**Abstract**

This paper is an investigation of the class of real classical Markov processes without a birth process that will generate the Dirac equation in 1+1 dimensions. The Markov process is assumed to evolve in an extra (ordinal) time dimension. The derivation requires that occupation by the Dirac particle of a space-time lattice site is encoded in a 4 state classical probability vector. Disregarding the state occupancy, the resulting Markov process is a homogeneous and almost isotropic binary random walk in Dirac space and Dirac time (including Dirac time reversals). It then emerges that the Dirac wavefunction can be identified with a polarization induced by the walk on the Dirac space-time lattice. The model predicts that QM observation must happen in ordinal time, and that wavefunction collapse is due not to a dynamical discontinuity, but to selection of a particular ordinal time. Consequently, the model is more relativistically equitable in its treatment of space and time in that the observer is attributed no special ability to specify the Dirac time of observation.

**Background**

There have been numerous recent attempts to derive the equations of Schrodinger and Dirac from a real, classical random process. In the works of Nelson [1], Nagasawa [2] El Naschie [3][4] and Nottale [5] a random walk is guided by a field in such a manner as to reproduce the required Schrodinger distribution. Such methods are close relatives of the pilot wave approach of De Broglie and Bohm which is clearly explicated by Holland [6]. Though they are a useful step in the direction of visualization, the paradoxical qualities of QM remain but in an altered form, namely the strange properties demanded of the guiding phenomenon. By contrast, formal analytic continuation motivated by Feynman's path integral approach [7] converts the diffusion equation into the Schrodinger equation. The merit of this approach is that the path integral method could be understood as a sum over all possible Brownian paths, but with complex weights. However, the technique is essentially mathematical rather than physical, and does not seem explicable in terms of classically visualizable processes. A major aspect of Ord's methods is to consider the wavefunction as linear difference of strictly classical occupation probabilities undergoing a Markov process [8][9][10]. In the opinion of the author, this seems to have been a crucial step forward. However, Ord's methods suffer from a lack of detailed balance of the probability, and therefore connote (i.e. implicitly) a particle birth process in time. Further, particle non-locality is achieved only if his probability is really that of an ensemble.

In common with Ord's work, the model presented herein constructs a Dirac wavefunction from probability differences. But it differs from his efforts by framing the Markov process in an 'ordinal' time other than the 'regular' (Dirac) time. This decoupling confers three particular advantages:
a)  QM non-locality arises naturally for a single particle, i.e. does not require an ensemble.
b)  The unphysical birth process required by Ord and others is replaced by a prior in ordinal time, and for which a physically reasonable explanation may be found.
c)  Wavefunction collapse may be understood as a lack of isomorphism between the two types of time.

The approach of this effort is first to present in some detail the steps by which the Dirac equation may be derived from a Markov random process. This is an ill-posed problem, so there is some commentary on the physical implications of the regularizing choices made. Subsequently, the random walk is illustrated, following which there is some discussion on the interpretation of the wavefunction and QM observation in terms of the conjectured underlying classical process.

**The Markov random process**

Let $p_i(t,x|\lambda)$ be the probability that the particle is in state $i$ and at $x$, $t$ at a given discrete ordinal-time $\lambda$. Let there be $n$ states, the occupation of which, at $x$, $t$, be denoted by a probability vector $\mathbf{p}(t,x|\lambda)$ of dimension $n$. That there is only one particle, which is always somewhere, requires that

$$\sum_{i=1}^{n} \int dt \int dx\, p_i(t,x|\lambda) = 1$$
$$i.e. \int dt \int dx\, \mathbf{1}.\mathbf{p}(t,x|\lambda) = 1 \tag{1}$$

Let the particle be subject to a scattering process at each of an incremental $\lambda$:

$$\mathbf{p}(t,x|\lambda+1) = \mathbf{T}(t,x)\mathbf{p}(t,x|\lambda) \tag{2}$$

where $\mathbf{T}$ is a transition matrix, which because of Eq. (2):

$$\forall \lambda : \int dt \int dx\, \mathbf{1}.\mathbf{p}(t,x|\lambda+1) = \int dt \int dx\, \mathbf{1}.\mathbf{T}(t,x)\mathbf{p}(t,x|\lambda) = 1 \tag{3}$$

and

$$\forall t,x : T_{i,j}(t,x) \in [0,1]. \tag{4}$$

**Ordinal-time average**

The aim is to construct a first order differential equation in $\mathbf{p}$ and then relate that to the Dirac wavefunction, for which it will be useful to construct an average over the ordinal-time $\lambda$:

$$\mathbf{p}(t,x) = \sum_{\lambda} \mathbf{p}(t,x|\lambda) p(\lambda) \tag{5}$$

which is the probability of the particle being at lattice site $x$, $t$ subject to whatever prior information there is about the ordinal time. Ultimately, in order to establish a connection between quantum-mechanical observation and the specification / selection of ordinal time, it will be necessary that the prior $p(\lambda)$ is exponential:

$$p(\lambda) = \left(\frac{1-\mu^{-1}}{1-\mu^{-\Lambda-1}}\right) \Theta(\Lambda-\lambda)\Theta(\lambda)\mu^{-\lambda}$$
$$= \left(\frac{1-\mu}{1-\mu^{1+\Lambda}}\right) \Theta(\Lambda-\lambda)\Theta(\lambda)\mu^{\Lambda-\lambda} \tag{6}$$

where $0 \leq \mu < 1$ and $\Theta$ is the Heavyside step function:

$$\Theta(x) = 1 \text{ if } x \geq 1$$
$$= 0 \text{ otherwise} \tag{7}$$

For large $\Lambda$, the prior is
$$p(\lambda) \approx (1-\mu)\mu^{\Lambda-\lambda}; \quad \lambda \in [0..\Lambda]$$
$$= 0; \quad \lambda \notin [0..\Lambda] \tag{8}$$

Eq. (8) is just an exponential decay away from the remote ordinal future into the ordinal past. The ordinal time can now be removed from the dynamics by multiplying Eq. (2) by $p(\lambda)$ and summing over $\lambda$ as follows

$$\sum_{\lambda=0}^{\Lambda}(1-\mu)\mu^{\Lambda-\lambda}\mathbf{p}(t,x\mid\lambda+1)=\sum_{\lambda=0}^{\Lambda}(1-\mu)\mu^{\Lambda-\lambda}\mathbf{T}(t,x)\mathbf{p}(t,x\mid\lambda)$$

$$\Rightarrow \sum_{\lambda=1}^{\Lambda+1}(1-\mu)\mu^{\Lambda-\lambda+1}\mathbf{p}(t,x\mid\lambda)=\mathbf{T}(t,x)\mathbf{p}(t,x) \qquad (9)$$

$$\Rightarrow -\mu\mathbf{p}(t,x)-(1-\mu)\big(\mathbf{p}(t,x\mid\Lambda+1)+\mu^{\Lambda+1}\mathbf{p}(t,x\mid 0)\big)=\mathbf{T}(t,x)\mathbf{p}(t,x)$$

$$\Rightarrow (\mathbf{T}(t,x)-\mu\mathbf{I})\mathbf{p}(t,x)\approx (1-\mu)\mathbf{p}(t,x\mid\Lambda+1)$$

We now impose the constraint that in the remote ordinal future the particle is well outside the space-time region of interest, $R_i$ say:

$$\mathbf{p}(t,x\mid\Lambda+1)=0 \quad \forall (t,x)\in R_i \qquad (10)$$

whereupon Eq. (9) reads

$$(\mathbf{T}(t,x)-\mu\mathbf{I})\mathbf{p}(t,x)=0 \quad \forall (t,x)\in R_i \,. \qquad (11)$$

This may be achieved by introducing a very slight time asymmetry in the transition matrix that has the effect of dragging the particle forward in time (with increasing ordinal time). Such a bias connotes an induced correlation between the ordinal time and Dirac time. The extent of $R_i$ for which Eq. (11) holds depends on the degree of time asymmetry in $\mathbf{T}$. The latter will, henceforth, be understood but not explicated here since the details are not relevant to the outcome of this effort.

**Relationship between Markov probability and the wavefunction**

As proposed by Ord [10], it will be assumed that the Dirac vector-wavefunction is a linear combination of state occupancy probabilities:

$$\mathbf{\psi}(t,x)=\mathbf{\Gamma}\mathbf{p}(t,x) \qquad (12)$$

where $\mathbf{\Gamma}$ is a real constant matrix. In 1+1 dimensions the Dirac $\mathbf{\psi}$ is a two-component vector. Each component may take both positive and negative values. Neglecting normalization (of both $\mathbf{p}$ and $\mathbf{\psi}$) each component of the probability vector does not furnish a full functional degree of freedom because each must be positive. Ord's major contribution was to establish a connection between quantum mechanics and a classical random walk by constructing a wavefunction from differences in the state occupancy probabilities. Though this may at first seem strange, it does have a straight-forward statistical and physical interpretation discussion of which there is some discussion toward the end of this document. It follows that $\mathbf{p}$ must have more than 2 states; Ord's assumption that there are 4 states is adopted here. Then, in order to deal with a square transformation matrix, it is convenient to append an extra two components into the definition of $\mathbf{\psi}$, the dynamics of which can be chosen without affecting the known physics of the Dirac wavefunction. Specifically, $\mathbf{\psi}$ will be taken to represent a 4-component vector, the first two components of which solve the Dirac equation in 1+1 dimensions, and $\mathbf{\Gamma}$ will be taken to represent a 4x4 matrix.

Ord's choice for $\Gamma$ assumes that the first two components of the vector wavefunction can be formed from probability differences:

$$\Gamma = \begin{pmatrix} \mathbf{I} & -\mathbf{I} \\ \mathbf{G} & \mathbf{G} \end{pmatrix}, \quad \Gamma^{-1} = \frac{1}{2}\begin{pmatrix} \mathbf{I} & \mathbf{G}^{-1} \\ -\mathbf{I} & \mathbf{G}^{-1} \end{pmatrix}. \tag{13}$$

It turns out that one does not have complete freedom to choose the dynamics of the extra two components of the wavefunction, and that **G** decides how those dynamics translate into the transition matrix **T**. Where it is useful to be explicit, we will adopt Ord's choice:

$$\mathbf{G} = \begin{pmatrix} 1 & -1 \\ 1 & 1 \end{pmatrix}, \quad \mathbf{G}^{-1} = \frac{1}{2}\begin{pmatrix} 1 & 1 \\ -1 & 1 \end{pmatrix}. \tag{14}$$

In the following a useful result will be that the similarity transform of a general block-diagonal matrix using this transformation is

$$\Gamma^{-1}\begin{pmatrix} \mathbf{S} & 0 \\ 0 & \mathbf{R} \end{pmatrix}\Gamma = \frac{1}{2}\begin{pmatrix} \mathbf{S} & -\mathbf{S} \\ -\mathbf{S} & \mathbf{S} \end{pmatrix} + \frac{1}{2}\begin{pmatrix} \tilde{\mathbf{R}} & \tilde{\mathbf{R}} \\ \tilde{\mathbf{R}} & \tilde{\mathbf{R}} \end{pmatrix} \tag{15}$$

where $\tilde{\mathbf{R}}$ is itself a similarity transform of **R**:

$$\tilde{\mathbf{R}} \equiv \mathbf{G}^{-1}\mathbf{R}\mathbf{G}. \tag{16}$$

I.E., introducing

$$\mathbf{H} \equiv \begin{pmatrix} 1 & -1 \\ -1 & 1 \end{pmatrix}, \tag{17}$$

Eq. (15) can be written compactly as

$$\Gamma^{-1}\begin{pmatrix} \mathbf{S} & 0 \\ 0 & \mathbf{R} \end{pmatrix}\Gamma = \frac{1}{2}\mathbf{H}\otimes\mathbf{S} + \frac{1}{2}\mathbf{I}\otimes\tilde{\mathbf{R}}. \tag{18}$$

To be explicit, the Dirac wavefunction in terms of the occupation probabilities of the 4 particle states, using Eq. (12) is

$$\begin{aligned}\psi_1(t,x) &= p_1(t,x) - p_3(t,x) \\ \psi_2(t,x) &= p_2(t,x) - p_4(t,x)\end{aligned} \tag{19}$$

and the extra two components are

$$\begin{aligned}\psi_2(t,x) &= p_1(t,x) - p_2(t,x) + p_3(t,x) - p_4(t,x) \\ \psi_4(t,x) &= p_1(t,x) + p_2(t,x) + p_3(t,x) + p_4(t,x)\end{aligned}. \tag{20}$$

**Extension to include jumps on the space-time lattice**

In order to arrive at a wave equation, scattering at each increment $\Delta\lambda$ is permitted not only between the states comprising the components of **p**, but also between adjacent cell-sites on the space-time lattice. Such behavior may be accommodated without increasing the dimensionality of **p** and **T** through the use of shift operators in space-time. Defining the primitive operator $\hat{E}_{j,k}(t,x)$,

$$\hat{E}_{j,k}(t,x)f(t,x) = f(t+j\Delta t, x+k\Delta x) \tag{21}$$

(where, without loss of generality, the space and time increments are made equal, $\Delta t = \Delta x$; $c = 1$) the transition matrix is assumed to have the form

$$\mathbf{T} = \hat{\mathbf{T}} = \mathbf{T}^{(0,0)} + \mathbf{T}^{(1,1)}\hat{E}_{1,1} + \mathbf{T}^{(-1,1)}\hat{E}_{-1,1} + \mathbf{T}^{(1,-1)}\hat{E}_{1,-1} + \mathbf{T}^{(-1,-1)}\hat{E}_{-1,-1}. \quad (22)$$

Following Feynman and Ord [10], only unit-diagonal hops along the space-time lattice, i.e. only particle jumps along the light-cone, are permitted. One could instead expand in terms of hops parallel to the space and time lattice axes (i.e. $\mathbf{T}^{(0,1)}$, $\mathbf{T}^{(1,0)}$, ...) which has the effect only of rotating the axes but does not change the dynamics.

Note that since the scattering coefficients for each shift operator must also be probabilities, the $T_{j,k}^{(a,b)}$ must satisfy

$$\forall a,b,j,k : T_{j,k}^{(a,b)} \in [0,1]. \quad (23)$$

Inserting Eq. (22) into Eq. (11) one obtains

$$\left(\mathbf{T}^{(0,0)} - \mu\mathbf{I} + \mathbf{T}^{(1,1)}\hat{E}_{1,1} + \mathbf{T}^{(-1,1)}\hat{E}_{-1,1} + \mathbf{T}^{(1,-1)}\hat{E}_{1,-1} + \mathbf{T}^{(-1,-1)}\hat{E}_{-1,-1}\right)\mathbf{p}(x,t) = \mathbf{0}. \quad (24)$$

Note that the Markov process framed as a random walk in ordinal time permits both forward and backward motion in Dirac time.

**Construction of a differential equation**

To construct a differential equation the shift operators must be expressed as differential operators. Since the random walk is essentially a diffusion process, the shift operator must be expanded at least to second order

$$\hat{E}_{j,k} \approx 1 + j\Delta x \frac{\partial}{\partial t} + k\Delta x \frac{\partial}{\partial x} + \frac{1}{2}(j\Delta x)^2 \frac{\partial^2}{\partial t^2} + jk(\Delta x)^2 \frac{\partial^2}{\partial x \partial t} + \frac{1}{2}(k\Delta x)^2 \frac{\partial^2}{\partial x^2}. \quad (25)$$

In the random walk framework the expression for $\mathbf{T}$ in terms of the $E_{j,k}$, and therefore Eq. (24), is exact, whilst the expansion of the $E_{j,k}$ in terms of first and second derivatives is an approximation. Eq. (25) is valid only if the higher-order terms are small by comparison; a condition that will have to be verified subsequently. An approximately sufficient condition is that the $p_i$ satisfy:

$$\frac{\Delta x}{2}\left|\frac{\partial^2 p_i}{\partial x^2}\right| << \left|\frac{\partial p_i}{\partial x}\right|, \quad \frac{\Delta x}{2}\left|\frac{\partial^2 p_i}{\partial t^2}\right| << \left|\frac{\partial p_i}{\partial t}\right|. \quad (26)$$

Considering that the increments are as yet undefined, this will be regarded as a constraint on $\Delta x$ and $\Delta t$, the smallest distance over which the $p_i$ are found to be relatively smoothly varying.

Using the expansion in Eq. (25), Eq. (11) is now

$$\left(\mathbf{T}^{(0,0)} + \mathbf{S}^{(\Sigma)} - \mu\mathbf{I} + \Delta x \mathbf{S}^{(t)} \frac{\partial}{\partial t} + \Delta x \mathbf{S}^{(x)} \frac{\partial}{\partial x} + \frac{1}{2}(\Delta x)^2 \mathbf{S}^{(\Sigma)} \frac{\partial^2}{\partial t^2} + \frac{1}{2}(\Delta x)^2 \mathbf{S}^{(\Sigma)} \frac{\partial^2}{\partial x^2}\right)\mathbf{p}(t,x) = \mathbf{0} \quad (27)$$

where the $\mathbf{S}^{(a)}$ are 'scattering' matrices satisfying

$$\mathbf{S}^{(\Sigma)} \equiv \mathbf{T}^{(1,1)} + \mathbf{T}^{(-1,1)} + \mathbf{T}^{(1,-1)} + \mathbf{T}^{(-1,-1)}$$
$$\mathbf{S}^{(t)} \equiv \mathbf{T}^{(1,1)} - \mathbf{T}^{(-1,1)} + \mathbf{T}^{(1,-1)} - \mathbf{T}^{(-1,-1)} \quad . \tag{28}$$
$$\mathbf{S}^{(x)} \equiv \mathbf{T}^{(1,1)} + \mathbf{T}^{(-1,1)} - \mathbf{T}^{(1,-1)} - \mathbf{T}^{(-1,-1)}$$

The normalization condition that conserves probability, Eq. (3) is now expressed as

$$\mathbf{1}.\left(\mathbf{T}^{(0,0)} + \mathbf{S}^{(\Sigma)}\right) = \mathbf{1} \quad . \tag{29}$$

Eq. (27) completes the development of the model in the forward direction (towards the Dirac wave-equation). It is now convenient to work backwards to this point from the Dirac equation with the aim of determining the matrices $\mathbf{S}^{(i)}$.

**Connection with Dirac's equation**

In a representation in which wavefunction is real there are two inequivalent 1+1 D Dirac equations which can be written:

$$\left(m - i\hbar\boldsymbol{\sigma}_y \frac{\partial}{\partial t} + \hbar\boldsymbol{\sigma}_x \frac{\partial}{\partial x}\right)\begin{pmatrix}\psi_1\\\psi_2\end{pmatrix} = 0 \tag{30}$$

and

$$\left(m - i\hbar\boldsymbol{\sigma}_y \frac{\partial}{\partial t} + \hbar\boldsymbol{\sigma}_z \frac{\partial}{\partial x}\right)\begin{pmatrix}\psi_1\\\psi_2\end{pmatrix} = 0 \tag{31}$$

where

$$\boldsymbol{\sigma}_x = \begin{pmatrix}0 & 1\\1 & 0\end{pmatrix}, \quad -i\boldsymbol{\sigma}_y = \begin{pmatrix}0 & -1\\1 & 0\end{pmatrix} = \boldsymbol{\sigma}_t \text{ say}, \quad \boldsymbol{\sigma}_z = \begin{pmatrix}1 & 0\\0 & -1\end{pmatrix}. \tag{32}$$

These two equations cannot related by a matrix factor or similarity transform. In this paper only the first will be investigated. Discussion of the second form is postponed to a presentation in 3+1 D. Also, it may be noted that either of these equations may be right-multiplied by a 2x2 matrix prior to making a connection with the Markov process. For instance, the scalar term may then be associated with the time derivative, rather than the mass. (This is the form adopted by Ord.) Such a transformation *does* has the effect of changing the master equations and therefore the geometry of the random walk. However, the mathematical procedure to arrive at the master equations – as described below - remains unchanged. A fuller discussion of the alternative geometries is postponed to a future report.

For the extended wavefunction let

$$\left(m\begin{pmatrix}\mathbf{I} & \mathbf{0}\\\mathbf{0} & \mathbf{R}^{(m)}\end{pmatrix} + \hbar\begin{pmatrix}\boldsymbol{\sigma}_t & \mathbf{0}\\\mathbf{0} & \mathbf{R}^{(t)}\end{pmatrix}\frac{\partial}{\partial t} + \hbar\begin{pmatrix}\boldsymbol{\sigma}_x & \mathbf{0}\\\mathbf{0} & \mathbf{R}^{(x)}\end{pmatrix}\frac{\partial}{\partial x} + \frac{\hbar^2}{m}\begin{pmatrix}\mathbf{0} & \mathbf{0}\\\mathbf{0} & \mathbf{R}^{(tt)}\end{pmatrix}\frac{\partial^2}{\partial t^2} + \frac{\hbar^2}{m}\begin{pmatrix}\mathbf{0} & \mathbf{0}\\\mathbf{0} & \mathbf{R}^{(xx)}\end{pmatrix}\frac{\partial^2}{\partial x^2}\right)\boldsymbol{\psi} = 0 \tag{33}$$

where $\boldsymbol{\psi} = (\psi_1, \psi_2, \psi_3, \psi_4)$. The block diagonal form of Eq. (33) uncouples the dynamics of dynamics of $\psi_1$, $\psi_2$ from those of $\psi_3$, $\psi_4$. Though the latter are considered to be outside the domain of 'observable physics', they have a status equal to that $\psi_1$, $\psi_2$ as elements of the random walk. Eq. (33) may now be transformed into the domain of the probability vector by using Eq. (12) and left-multiplying by $\boldsymbol{\Gamma}^{-1}$. Equating the so-transformed Dirac equation with Eq. (27) gives:

$$\mathbf{T}^{(0,0)} + \mathbf{S}^{(\Sigma)} = \mu\boldsymbol{\Gamma}^{-1}\begin{pmatrix}\mathbf{0} & \mathbf{0}\\\mathbf{0} & \mathbf{I} - \mathbf{R}^{(m)}\end{pmatrix}\boldsymbol{\Gamma} \tag{34}$$

$$\mathbf{S}^{(t)} = -\kappa\mu\mathbf{\Gamma}^{-1}\begin{pmatrix} \sigma_t & 0 \\ 0 & \mathbf{R}^{(t)} \end{pmatrix}\mathbf{\Gamma} \qquad (35)$$

$$\mathbf{S}^{(x)} = -\kappa\mu\mathbf{\Gamma}^{-1}\begin{pmatrix} \sigma_x & 0 \\ 0 & \mathbf{R}^{(x)} \end{pmatrix}\mathbf{\Gamma} \qquad (36)$$

$$\mathbf{S}^{(\Sigma)} = -2\kappa^2\mu\mathbf{\Gamma}^{-1}\begin{pmatrix} 0 & 0 \\ 0 & \mathbf{R}^{(tt)} \end{pmatrix}\mathbf{\Gamma} = -2\kappa^2\mu\mathbf{\Gamma}^{-1}\begin{pmatrix} 0 & 0 \\ 0 & \mathbf{R}^{(xx)} \end{pmatrix}\mathbf{\Gamma} \qquad (37)$$

where
$$\kappa \equiv \frac{\hbar}{m\Delta x} \qquad (38)$$

for which the inequality Eq. (26) requires that

$$\kappa \gg \frac{\hbar k}{2m}, \quad \kappa \gg \frac{\hbar\omega}{2m} \qquad (39)$$

where $\omega$, $k$ are the traditional (quantum-mechanical) particle frequency and wave-vector. Performing the similarity transforms using Eq. (15) gives the following set of equations:

$$\mathbf{T}^{(0,0)} + \mathbf{S}^{(\Sigma)} = \frac{\mu}{2}\begin{pmatrix} \mathbf{I} - \widetilde{\mathbf{R}}^{(m)} & \mathbf{I} - \widetilde{\mathbf{R}}^{(m)} \\ \mathbf{I} - \widetilde{\mathbf{R}}^{(m)} & \mathbf{I} - \widetilde{\mathbf{R}}^{(m)} \end{pmatrix} \qquad (40)$$

where, due to Eq. (29)

$$\mathbf{1}.\widetilde{\mathbf{R}}^{(m)} = \frac{\mu - 1}{\mu} \qquad (41)$$

i.e. Eq. (40) can be written

$$\mathbf{T}^{(0,0)} + \mathbf{S}^{(\Sigma)} = \frac{1}{2}\mathbf{I} \otimes \begin{pmatrix} c_1 & 1 - c_2 \\ 1 - c_1 & c_2 \end{pmatrix} \qquad (42)$$

where now the degrees of freedom are encapsulated in the $c_i$, and

$$\widetilde{\mathbf{R}}^{(m)} = \frac{1}{\mu}\begin{pmatrix} \mu - c_1 & c_2 - 1 \\ c_1 - 1 & \mu - c_2 \end{pmatrix}. \qquad (43)$$

Equation (35) is

$$\mathbf{S}^{(t)} = -\frac{\kappa\mu}{2}\begin{pmatrix} \widetilde{\mathbf{R}}^{(t)} + \sigma_t & \widetilde{\mathbf{R}}^{(t)} - \sigma_t \\ \widetilde{\mathbf{R}}^{(t)} - \sigma_t & \widetilde{\mathbf{R}}^{(t)} + \sigma_t \end{pmatrix}, \qquad (44)$$

Eq. (36) is

$$\mathbf{S}^{(x)} = -\frac{\kappa\mu}{2}\begin{pmatrix} \widetilde{\mathbf{R}}^{(x)} + \sigma_x & \widetilde{\mathbf{R}}^{(x)} - \sigma_x \\ \widetilde{\mathbf{R}}^{(x)} - \sigma_x & \widetilde{\mathbf{R}}^{(x)} + \sigma_x \end{pmatrix}, \qquad (45)$$

and Eq. (37) gives both

$$\mathbf{S}^{(\Sigma)} = -\kappa\mu\begin{pmatrix} \widetilde{\mathbf{R}}^{(tt)} & \widetilde{\mathbf{R}}^{(tt)} \\ \widetilde{\mathbf{R}}^{(tt)} & \widetilde{\mathbf{R}}^{(tt)} \end{pmatrix} \qquad (46)$$

and

$$\widetilde{\mathbf{R}}^{(xx)} = \widetilde{\mathbf{R}}^{(tt)} \Rightarrow \mathbf{R}^{(xx)} = \mathbf{R}^{(tt)}. \qquad (47)$$

The goal is to solve for the $\mathbf{S}^{(i)}$ from Eqs. (42), (44), (45), and (46), and then solve for the $\mathbf{T}^{(i,j)}$ from the $\mathbf{S}^{(i)}$ from Eqs. (28) and (29). Clearly the $\mathbf{S}^{(i)}$, and therefore the $\mathbf{T}^{(i,j)}$ are under-constrained by this system of equations. The implication is that there is a family of random walks from which the Dirac equation may be extracted. In the following, where possible, the extra degrees of freedom will be pruned by appealing to the underlying physics. However: it is not claimed that the choices are inevitable or unique.

**Separation of drift and diffusion**

From Eq. (33) the $\psi_3$, and $\psi_4$ satisfy

$$\left( m\mathbf{R}^{(m)} + \hbar\mathbf{R}^{(t)}\frac{\partial}{\partial t} + \hbar\mathbf{R}^{(x)}\frac{\partial}{\partial x} + \frac{\hbar^2}{m}\mathbf{R}^{(tt)}\left(\frac{\partial^2}{\partial t^2} + \frac{\partial^2}{\partial x^2}\right)\right)\begin{pmatrix}\psi_3\\ \psi_4\end{pmatrix} = 0 \;. \tag{48}$$

Although the equations for the 'observable' wavefunction $\psi_1$, and $\psi_2$ are decoupled from those in $\psi_3$, and $\psi_4$, there are constraints on the latter. Note that $\mathbf{R}^{(tt)}$ cannot be zero, because Eqs. (46) and (28) would then imply that $\mathbf{S}^{(t)}$ and $\mathbf{S}^{(x)}$ were also zero, making Eqs. (44) and (45) an impossible condition on $\mathbf{R}^{(t)}$ and $\mathbf{R}^{(x)}$. Therefore the last two terms in the above are mandatory, just as their absence in the equation for $\psi_1$, and $\psi_2$ is mandatory.

Similarly, the first three terms are mandatory in the Dirac operator on $\psi_1$ and $\psi_2$. But they are optional in the equation for $\psi_3$, and $\psi_4$. Broadly speaking: the Dirac equation is a propagation equation; the equation in $\psi_3$, and $\psi_4$ is a diffusion equation (in 1+1 D, taking place in particle time). The appearance of $\mathbf{R}^{(t)}$ and $\mathbf{R}^{(x)}$ in the latter has the character of drift – akin to the propagation behavior of $\psi_1$ and $\psi_2$. Therefore, one may regard the Dirac equation as parsing out the drift aspect of pair-wise differences of diffusion processes. In order to make a clean split, and in the interests of keeping the model as simple as possible whilst consistent with known physics, the equation in $\psi_3$, and $\psi_4$ will be kept entirely free of drift, so that it describes only a diffusion behavior. This may be achieved by setting

$$\widetilde{\mathbf{R}}^{(t)} = \widetilde{\mathbf{R}}^{(x)} = \mathbf{R}^{(t)} = \mathbf{R}^{(x)} = 0$$

whereupon Eq. (48) becomes

$$\left( m^2\mathbf{R}^{(m)} + \hbar^2\mathbf{R}^{(tt)}\left(\frac{\partial^2}{\partial t^2} + \frac{\partial^2}{\partial x^2}\right)\right)\begin{pmatrix}\psi_3\\ \psi_4\end{pmatrix} = 0 \;. \tag{49}$$

This leaves a simplified set of equations for the transition matrices. From Eqs. (28) and (42):

$$\mathbf{T}^{(0,0)} + \mathbf{T}^{(1,1)} + \mathbf{T}^{(-1,1)} + \mathbf{T}^{(1,-1)} + \mathbf{T}^{(-1,-1)} = \frac{1}{2}\mathbf{I}\otimes\begin{pmatrix} c_1 & 1-c_2 \\ 1-c_1 & c_2 \end{pmatrix}, \tag{50}$$

Eq. (44) becomes

$$\mathbf{T}^{(1,1)} - \mathbf{T}^{(-1,1)} + \mathbf{T}^{(1,-1)} - \mathbf{T}^{(-1,-1)} = -\frac{\kappa\mu}{2}\mathbf{H}\otimes\boldsymbol{\sigma}_t, \tag{51}$$

Eq. (45) becomes

$$\mathbf{T}^{(1,1)} + \mathbf{T}^{(-1,1)} - \mathbf{T}^{(1,-1)} - \mathbf{T}^{(-1,-1)} = -\frac{\kappa\mu}{2}\mathbf{H}\otimes\boldsymbol{\sigma}_x, \tag{52}$$

and Eq. (46) is

$$\mathbf{T}^{(1,1)} + \mathbf{T}^{(-1,1)} + \mathbf{T}^{(1,-1)} + \mathbf{T}^{(-1,-1)} = -\kappa\mu \mathbf{I} \otimes \widetilde{\mathbf{R}}^{(tt)}. \tag{53}$$

**No internal transitions**

The system of Eqs. (50)-(53) is still under-constrained; for instance it is apparent that $\mathbf{T}^{(0,0)}$ may be set to zero. $\mathbf{T}^{(0,0)}$ governs the transitions between states without movement on the lattice. With no internal transitions the particle may only change state when there is movement on the space-time lattice. It might be said that the particle is then structureless. Removal of $\mathbf{T}^{(0,0)}$ is in accord with the aim of constructing the simplest possible random walk out of which the Dirac equation can be projected.

**Sparse total transition matrix**

In solving for the $\mathbf{T}^{(a,b)}$, let the *total* transition matrix be as sparse as possible. Such a constraint minimizes the number of possible directions of particle motion out of each point on the space-time lattice (in this case it will be 2, i.e. a binary random walk). One approach is try to ensure that no element of the transition matrices can be simultaneously non-zero in more than one matrix. I.E.:

$$\forall j,k; \forall a \neq c, b \neq d : T_{j,k}^{(a,b)} T_{j,k}^{(c,d)} = 0. \tag{54}$$

Then, from Eqs. (51) and (52) one finds that the $\mathbf{T}^{(a,b)}$ must be traceless, and it follows from Eq. (50) that

$$c_1 = c_2 = 0 \tag{55}$$

from Eq. (43) this means

$$\widetilde{\mathbf{R}}^{(m)} = \begin{pmatrix} 1 & -1/\mu \\ -1/\mu & 1 \end{pmatrix} \tag{56}$$

with Eq. (14) and (16) this gives

$$\mathbf{R}^{(m)} = \begin{pmatrix} 1 + 1/\mu & 0 \\ 0 & 1 - 1/\mu \end{pmatrix} \tag{57}$$

and now Eqs. (50)-(53) with $\mathbf{T}^{(0,0)} = 0$ give

$$\widetilde{\mathbf{R}}^{(tt)} = -\frac{1}{2\kappa\mu} \sigma_x. \tag{58}$$

and

$$\mathbf{T}^{(1,1)} + \mathbf{T}^{(-1,1)} + \mathbf{T}^{(1,-1)} + \mathbf{T}^{(-1,-1)} = \frac{1}{2} \mathbf{I} \otimes \sigma_x, \tag{59}$$

$$\mathbf{T}^{(1,1)} - \mathbf{T}^{(-1,1)} + \mathbf{T}^{(1,-1)} - \mathbf{T}^{(-1,-1)} = -\frac{\kappa\mu}{2} \mathbf{H} \otimes \sigma_t, \tag{60}$$

$$\mathbf{T}^{(1,1)} + \mathbf{T}^{(-1,1)} - \mathbf{T}^{(1,-1)} - \mathbf{T}^{(-1,-1)} = -\frac{\kappa\mu}{2} \mathbf{H} \otimes \sigma_x. \tag{61}$$

It may readily be inferred from Eqs. (59)-(61) that the constraint Eq. (54) is possible only if

$$\kappa\mu = 1 \tag{62}$$

which, from Eq. (38), and restoring $c$, gives

$$\Delta x = \mu \frac{\hbar}{mc}, \quad \Delta t = \mu \frac{\hbar}{mc^2}. \tag{63}$$

**Solution for the transition matrices**

From Eqs. (59)-(62) the transition matrices may be solved using uniquely to give

$$\mathbf{T}^{(1,-1)} = \frac{1}{2}\begin{pmatrix} 0 & 1 & 0 & 0 \\ 0 & 0 & 0 & 0 \\ 0 & 0 & 0 & 1 \\ 0 & 0 & 0 & 0 \end{pmatrix}, \mathbf{T}^{(-1,1)} = \frac{1}{2}\begin{pmatrix} 0 & 0 & 0 & 1 \\ 0 & 0 & 0 & 0 \\ 0 & 1 & 0 & 0 \\ 0 & 0 & 0 & 0 \end{pmatrix}, \mathbf{T}^{(1,1)} = \frac{1}{2}\begin{pmatrix} 0 & 0 & 0 & 0 \\ 0 & 0 & 1 & 0 \\ 0 & 0 & 0 & 0 \\ 1 & 0 & 0 & 0 \end{pmatrix}, \mathbf{T}^{(-1,-1)} = \frac{1}{2}\begin{pmatrix} 0 & 0 & 0 & 0 \\ 1 & 0 & 0 & 0 \\ 0 & 0 & 0 & 0 \\ 0 & 0 & 1 & 0 \end{pmatrix}. \tag{64}$$

The total transition matrix operator, from Eq. (22), is therefore

$$\mathbf{T}(t,x) = \frac{1}{2}\begin{pmatrix} 0 & \hat{E}_{1,-1} & 0 & \hat{E}_{-1,1} \\ \hat{E}_{-1,-1} & 0 & \hat{E}_{1,1} & 0 \\ 0 & \hat{E}_{-1,1} & 0 & \hat{E}_{1,-1} \\ \hat{E}_{1,1} & 0 & \hat{E}_{-1,-1} & 0 \end{pmatrix} \tag{65}$$

and, with reference to Eq. (2), the master equations for the random walk are

$$\begin{aligned} p_1(t,x \mid \lambda+1) &= \frac{1}{2}\left[p_2(t+\Delta t, x-\Delta x \mid \lambda) + p_4(t-\Delta t, x+\Delta x \mid \lambda)\right] \\ p_2(t,x \mid \lambda+1) &= \frac{1}{2}\left[p_1(t-\Delta t, x-\Delta x \mid \lambda) + p_3(t+\Delta t, x+\Delta x \mid \lambda)\right] \\ p_3(t,x \mid \lambda+1) &= \frac{1}{2}\left[p_2(t-\Delta t, x+\Delta x \mid \lambda) + p_4(t+\Delta t, x-\Delta x \mid \lambda)\right] \\ p_4(t,x \mid \lambda+1) &= \frac{1}{2}\left[p_1(t+\Delta t, x+\Delta x \mid \lambda) + p_3(t-\Delta t, x-\Delta x \mid \lambda)\right] \end{aligned} \tag{66}$$

**Validity of the differential expansion of the shift operators**

It is important to validate the expansion of the shift operators given in Eq. (25). To this end, using the result for the total transition matrix, the shift-operator version of the Dirac equation is compared with the differential form. Using Eqs. (11), (12) and Eq. (65), the exact equation of motion for the Dirac wavefunction from the perspective of the random walk is

$$\begin{pmatrix} 1 & \frac{1}{2\mu}\left(\hat{E}_{-1,1} - \hat{E}_{1,-1}\right) \\ \frac{1}{2\mu}\left(\hat{E}_{1,1} - \hat{E}_{-1,-1}\right) & 1 \end{pmatrix}\begin{pmatrix} \psi_1 \\ \psi_2 \end{pmatrix} = 0. \tag{67}$$

By contrast, the 'canonical' Dirac equation of motion (equivalent to expanding the shift operators to second order), using Eq. (25), is

$$\begin{pmatrix} 1 & \frac{\hbar}{mc}\left(\frac{\partial}{\partial x} - \frac{\partial}{c\partial t}\right) \\ \frac{\hbar}{mc}\left(\frac{\partial}{\partial x} + \frac{\partial}{c\partial t}\right) & 1 \end{pmatrix}\begin{pmatrix} \psi_1 \\ \psi_2 \end{pmatrix} = 0. \tag{68}$$

For Eq. (68) to be a valid approximation to Eq. (67), it can be seen that the wavefunction must be relatively smooth along the light-cone. Specifically, $\psi_1$ must change relatively little for double hops along the line $x + ct = 0$, and $\psi_2$ must change relatively little for double hops along $x - ct = 0$. To be more precise it is convenient to compare the two equations in the Fourier domain. Defining

$$\tilde{\psi}(\omega, k) = \int dt \int dx \, e^{-i(\omega t + kx)} \psi(t, x) \tag{69}$$

Eq. (67) becomes

$$\begin{pmatrix} 1 & \frac{i}{\mu} \sin(k\Delta x - \omega \Delta t) \\ \frac{i}{\mu} \sin(k\Delta x + \omega \Delta t) & 1 \end{pmatrix} \begin{pmatrix} \tilde{\psi}_1 \\ \tilde{\psi}_2 \end{pmatrix} = 0. \tag{70}$$

Setting the determinant to zero and using Eq. (63) gives

$$\mu^2 + \sin\left(\mu \frac{\hbar}{mc}(k - \omega/c)\right) \sin\left(\mu \frac{\hbar}{mc}(k + \omega/c)\right) = 0. \tag{71}$$

which must be compared to the Einstein relation resulting from the same procedure applied to Eq. (68):

$$(\hbar \omega)^2 = (\hbar k c)^2 + m^2 c^4. \tag{72}$$

For equivalence of Eqs. (71) and (72) it can be seen that the arguments of both trig functions must be small, which can be satisfied only if

$$\mu \frac{\hbar k}{mc}, \mu \frac{\hbar \omega}{mc^2} << 1. \tag{73}$$

From Eq. (8), the exponential (ordinal) lifetime of the particle is $-1/\log(\mu)$. Since $\mu\kappa$ has been defined to be a constant in Eq. (62), $\mu$ also specifies the lattice jump-length as a fraction of the Compton wavelength $\lambda_c$:

$$\Delta x = \frac{\mu}{2\pi} \lambda_c. \tag{74}$$

For the random walk version of the Dirac equation to be valid up to around the Compton frequency, Eq. (73) demands that $\mu$ is a small number, independent of the mass of the particle. $\mu$ cannot be zero however, because then the Einstein relation would be lost from Eq. (71). Also, given that the Dirac equation is invalid beyond the Compton frequency, further agreement with the physics would flow from the random walk model if $\mu$ was not allowed to become very small.

**Description of the random walk**

The master Eqs. (66) generate the state transitions on the lattice illustrated in Fig. 1, where it is to be understood that the particle is always progressing monotonically in its own ordinal time. (The figure is similar to that used by Maxwell [11] to explain his luminiferous aether.) Several observations are worthy of emphasis:

- Space-time motion is rectilinear for a fixed interval along the light-cone.
- After each such space-time interval, the motion is changed with equal probability to one of the two orthogonal directions.
- Disregarding the state labels, the motion is an isotropic binary random walk in space and Dirac time.

- Without internal transitions (see above discussion concerning $\mathbf{T}^{(0,0)}$) each lattice site may accommodate only one of the four state labels.
- It follows that the state labels may be attributed to the lattice rather than the particle.
- If so regarded, then the 1+1 D wavefunction is a projection of a conventional binary random walk in space and time onto four sub-spaces.
- The 'spin' of the particle is not an intrinsic property of the particle, but is encoded in the wavefunction via the projection.
- It is evident that the random walk may be considered as resulting from a space-time generalization of a random magnetic field acting on a classical 'charge' in the sense that the orientation of motion may change, but the speed is unaffected.

**The wavefunction**

Let the 4 probability states be considered as a direct product of two binary-valued 'quantum numbers' $j,s$:

$$\begin{aligned} \mathbf{p}(t,x|\lambda) &\equiv \{p_1(t,x|\lambda), p_2(t,x|\lambda), p_3(t,x|\lambda), p_4(t,x|\lambda)\} \\ &= \{\pi_1(t,x,1|\lambda), \pi_2(t,x,1|\lambda), \pi_1(t,x,-1|\lambda), \pi_2(t,x,-1|\lambda)\}. \\ &= \{\pi_j(t,x,s|\lambda)\}; \quad j=1,2 \quad s=1,-1 \end{aligned} \quad (75)$$

With this choice of labeling, the Dirac wavefunction may be written

$$\psi_j(t,x) = \sum_{s=-1,1} s \int d\lambda \, \pi_j(t,x,s|\lambda) p(\lambda) = \sum_{s=-1,1} s \pi_j(t,x,s) = \langle s_j(t,x) \rangle; \quad j=1,2. \quad (76)$$

The wavefunction is the expected value of the $s$ quantum number, i.e. the wavefunction codes for a polarization $s_j(t,x)$ induced on the lattice.

**Dirac and random walk currents**

The classical equation of current-conservation in 1+1 D is

$$\frac{\partial \rho(t,x)}{\partial t} + \frac{\partial j(t,x)}{\partial x} = 0 \quad (77)$$

from which one readily infers

$$\frac{\partial}{\partial t} \int dx \, \rho(t,x) = 0, \quad \frac{\partial}{\partial x} \int dt \, j(t,x) = 0. \quad (78)$$

Eq. (77) can be constructed from the Dirac wavefunction provided one uses

$$\begin{aligned} \rho(t,x) &= \psi_1^2(t,x) + \psi_2^2(t,x) \\ j(t,x) &= \psi_2^2(t,x) - \psi_1^2(t,x) \end{aligned} \quad (79)$$

as may be verified by substitution of these expressions into Eq. (30). Therefore, in QM, the expressions for $\rho$ and $j$ in the above are associated with the classical charge and current respectively. (The sign of $\rho$ is fixed once established, say at $t = 0$, but is otherwise arbitrary.) Charge conservation so defined is a property of the Dirac equation, and therefore must be true of the random walk. It follows that the quantity

$$\rho(t,x) = \langle s_1(t,x)\rangle^2 + \langle s_2(t,x)\rangle^2 \qquad (80)$$

is the random walk interpretation of the charge density: the square of the expected polarization is a density whose integral over space does not vary in time.

**Exponential prior**

Ord (e.g. [10]) has achieved wave-like behavior from a diffusion process by implicitly introducing a birth process at each scattering in Dirac time. The effort reported here is distinguished by a motion in ordinal time without a birth process in Dirac time. Even so, the exponential prior $p(\lambda)$ might be construed as a birth process in $\lambda$ unless there is some reason to view it otherwise. But having started out with the idea of a conserved particle underlying the Dirac wavefunction, any birth process is unattractive and unphysical. Note that if a birth process is accepted, the explanation for wave-function collapse proffered below would not work. This is because one would no longer be able to associate a definite position with a particular ordinal time (selection), since there would be many particles occupying the lattice at every ordinal moment. (It follows that, because the weighting does not enter into the dynamics, in a numerical simulation the random walk may be executed without the prior, which need be invoked only upon observation. By contrast, a literal numerical simulation of the birth process requires an ensemble of particles evolving in parallel.) Presently however, a satisfactory physical explanation for the exponential prior is lacking.

**Wavefunction collapse**

Canonically (i.e. in QM), wavefunction collapse is short hand for the apparent tendency of the Dirac particle to occupy a definite position upon observation, whilst the dynamics of the wavefunction connotes simultaneous occupation of a continuum of positions in the manner of a wave rather than a particle. (Bohm's position is implicitly adopted here: that all observation is ultimately of position.)

A property of the exponential prior is that in any space-time volume there is only one space-time coordinate pair available for observation. This is (approximately) the space-time location of the last ordinal visit to that volume. (A slight forward time drift is enough to break from the traditional random walk rule that there will always be another visit in the ordinal future.) The same applies to a volume which is a very thin time slice, from which one deduces that the particle does have a definite position at every Dirac time, namely the position with the latest ordinal time stamp in that slice, as illustrated in Fig. 2. This would seem to be in contradiction with the canonical interpretation of the wavefunction. However, from the perspective of the random walk in ordinal time, the observer cannot specify the Dirac time, but instead supplies an ordinal time. According to this model, the relationship between the two is only statistical. It is the lack of a deterministic isomorphism between the two that gives the impression that the particle is occupying multiple positions at a single time.

**Summary and further work**

The model presented succeeds in reproducing the free-particle Dirac equation in 1+1 D from a classical random walk. It offers the possibility of a better understanding of the nature of the wavefunction, and the paradox of wavefunction collapse. The model can easily be extended to 3+1 dimensions, which will be presented subsequently. A satisfactory physical explanation for the ordinal-time exponential prior is lacking.

**Acknowledgments**

Harold Puthoff introduced me to Stochastic Electrodynamics as a means of understanding how QM behavior can derive from classical systems driven by noise. Garnet Ord showed me how to connect the QM wavefunction to a Markov distribution. I am immeasurably indebted and grateful to both of these people for countless hours of most enjoyable discussion, for their generous sharing of their time, and for patient explanations of their considerable understanding of their respective fields.

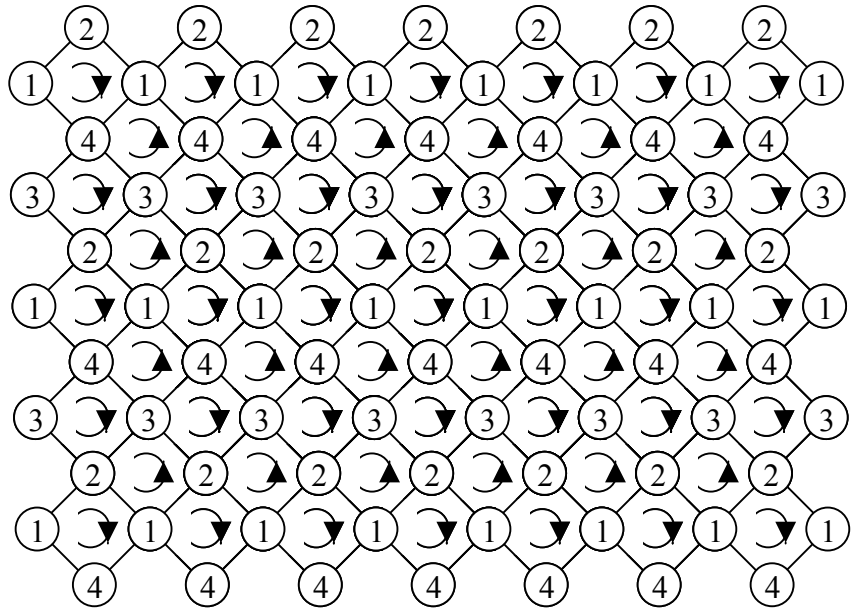 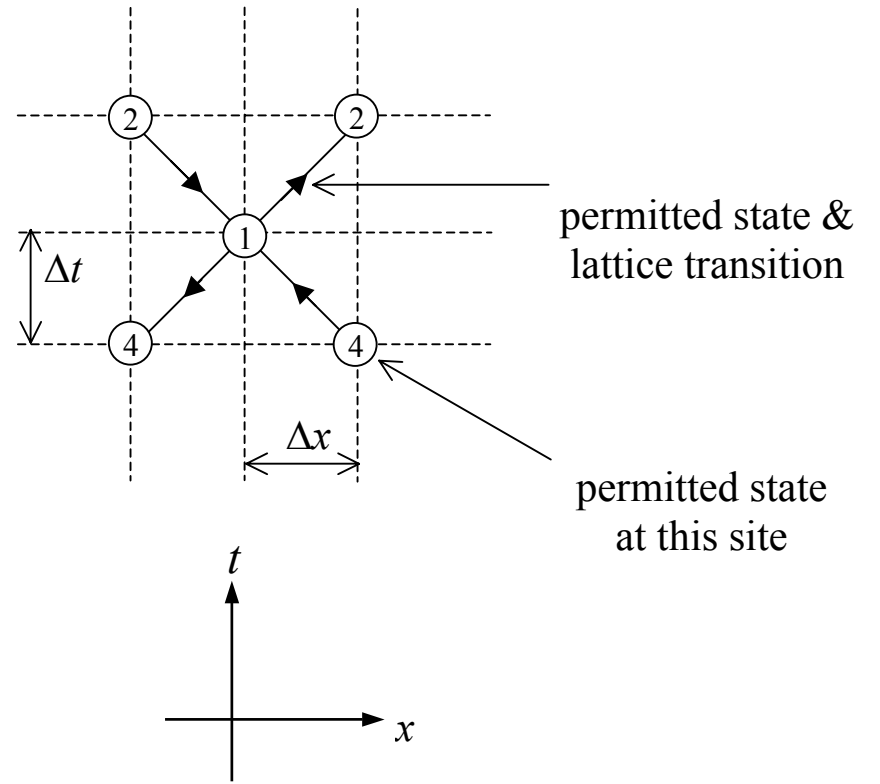

**Figure 1: State transitions on the lattice**

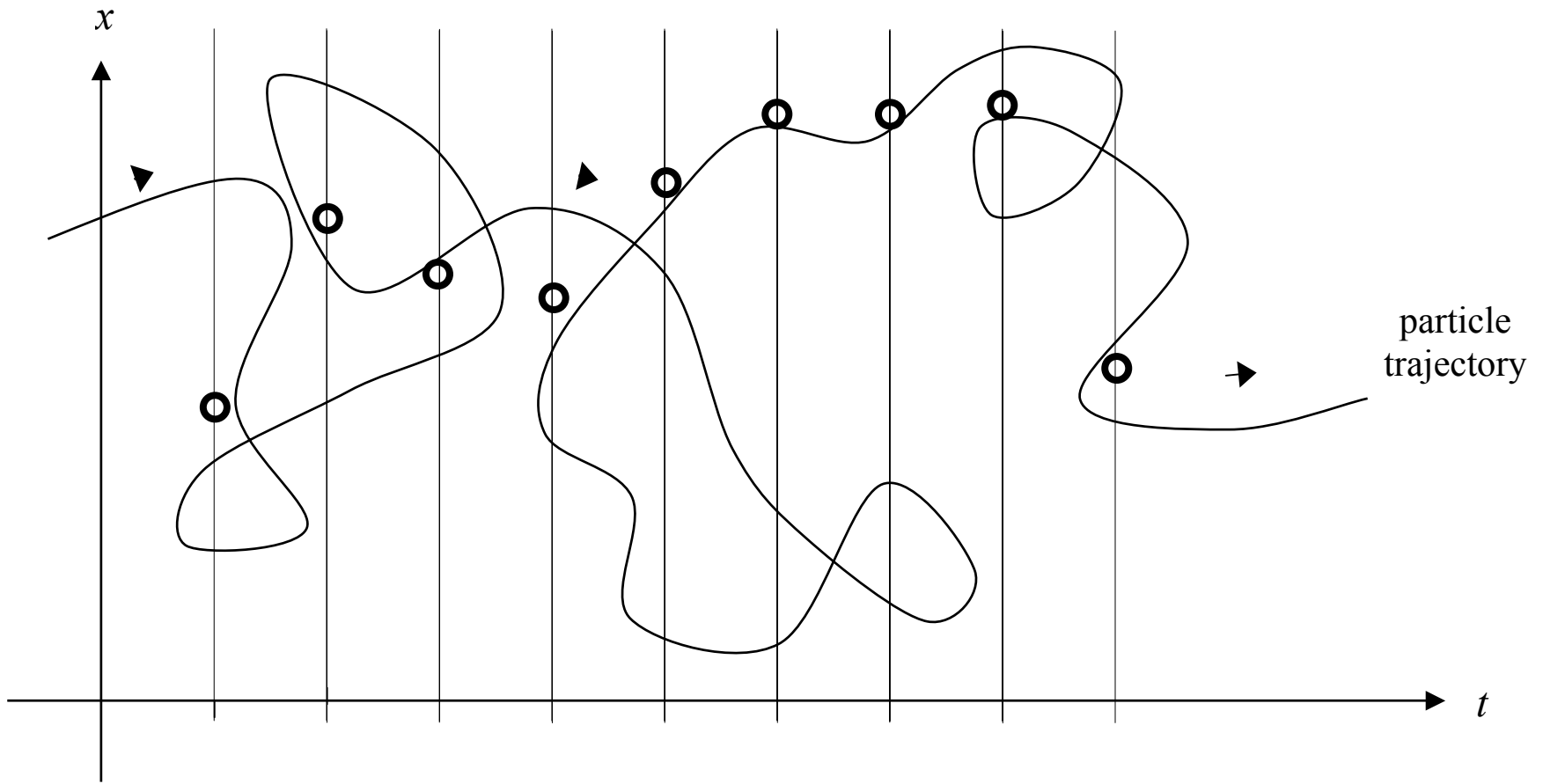

**Figure 2: example trajectory**

The circles mark the coordinates of the latest ordinal time visit to each Dirac time. These are the unique values of *x* that would be observed if the Dirac time could be specified.